\documentclass[11pt]{article}
\usepackage{graphicx}
\usepackage{hyperref}
\usepackage{tikz}
\usetikzlibrary{arrows}

\title{Emerge of scaling in project schedules}
\author{Alexei Vazquez\\
\\
\normalsize{Nodes \& Links Ltd, Salisbury House, Station Road, Cambridge, CB1 2LA, UK}
\\
\normalsize{$^\ast$To whom correspondence should be addressed; E-mail: alexei@nodeslinks.com}
}

\date{}

\begin{document}

\maketitle

\begin{abstract}
A project schedule contains a network of activities, the activity durations, the early and late finish dates for each activity, and the associated total float or slack times, the difference between the late and early dates. Here I show that the distribution of activity durations and total floats of construction project schedules exhibit a power law scaling. The power law scaling of the activity durations is explained by a historical process of specialization fragmenting old activities into new activities with shorter duration. In contrast, the power law scaling of the total floats distribution across activities is determined by the activity network. I demonstrate that the power law scaling of the activity duration distribution is essential to obtain a good estimate of the project delay distribution, while the actual total float distribution is less relevant. Finally, using extreme value theory and scaling arguments, I provide a mathematical proof for reference class forecasting for the project delay distribution. The project delay cumulative distribution function is  $G(z) = \exp( - (z_c/z)^{1/s})$, where $s>0$ and $z_c>0$ are shape and scale parameters. Furthermore, if activity delays follow a lognormal distribution, as the empirical data suggests, then $s=1$ and $z_c \sim N^{0.20}d_{\max}^{1+0.20(1-\gamma_d)}$, where $N$ is the number of activities, $d_{\max}$ the maximum activity duration in units of days and $\gamma_d$ the power law exponent of the activity duration distribution. These results offer new insights about project schedules, reference class forecasting and delay risk analysis.
\end{abstract}

\section{Introduction}

A project is a set of activities leading to a common goal. The activities are characterized by their durations and dependencies. The duration indicates how long it takes to finish a discrete activity. The dependencies list all activities that must finish before an activity can start. For example, consider a toy project with 3 activities A, B and C, with durations of 10, 2 and 4 days, respectively, and the dependency that B most finish before C can start
\begin{center}
\begin{tikzpicture}[->,>=stealth',auto,node distance=1.5cm,
	thick,main node/.style={font=\sffamily\small}]
\node[main node] (1) {start};
\node[main node] (2) [above right of=1] {AAAAAAAAAA};
\node[main node] (3) [below right of=1] {BB};
\node[main node] (4) [right of=3] {CCCC};
\node[main node] (5) [above right of=4] {end};
\path[every node/.style={font=\sffamily\small}]
	(1) 	edge [red] (2)
		edge  (3)	
	(3) edge (4)
	(2) edge [red] (5)
	(4) edge (5);
\end{tikzpicture}
\label{example}
\end{center}
The project ends when both A and C have finished. The duration of A (10 days) is larger than the total duration of B plus C (6 days) and therefore A sets the project duration. We say A is a critical activity \cite{pmbokguide2010}. In general project schedules has a critical path of activities that set the project duration. In turn, there is a float or slack time of 4 days between finishing B or C and finishing the project. We say B and C have a total float of 4 days. We can delay the finish of either B or C by 4 days without affecting the project end date. That is the basics of project scheduling \cite{van13}.

We have a good understanding of the statistics of activity dependencies \cite{ellinas19, santolini21, pozzana21, vazquez23}. The activity dependencies form an activity network where nodes are activities and an arrow between two activities indicates that the source must finish before the target starts. The number of direct predecessors to a reference activity are all other activities with an arrow pointing to the reference activity. Also called the in-degree. The number of direct successors to a reference activity are all other activities with an arrow from the reference activity. Also called the out-degree. Based on data for construction project schedules, the distribution of in- and out-degrees across activities are the same and they are both characterized by a power law tail $P(k)\sim k^{-\gamma_k}$, where $k$ denotes either the in-degree or the out-degree and $\gamma_k$ is the associated power-law exponent \cite{vazquez23}. The exponent $\gamma_k$ is project dependent and it tesll how heterogenous is the distribution of dependencies.

The power law scaling of the in- and out-degree distributions is explained by a process of specialization \cite{vazquez23}. For example, at some point in time there was an activity called building the foundations that "splits" into excavating and  poring-concrete, with the dependency that excavating must finish before poring-concrete starts. In another scenario, at some point in time there was an activity called interior-work that "duplicated" into electrical-work and plumbing, whereby both electrical-work and plumbing inherited all the backward and forward dependancies of interior-work. This duplication-split model recapitulates the power law scaling $P(k)\sim k^{-\gamma_k}$ with $\gamma_k=1/q$, where $q$ is the quotient of duplication per specialization event.

What is less known is what is the statistics of activity durations and total floats and what is their origin. Here I address these questions using data from construction projects and the duplication-split model. 

\begin{figure}[t]
\includegraphics[width=6.3in]{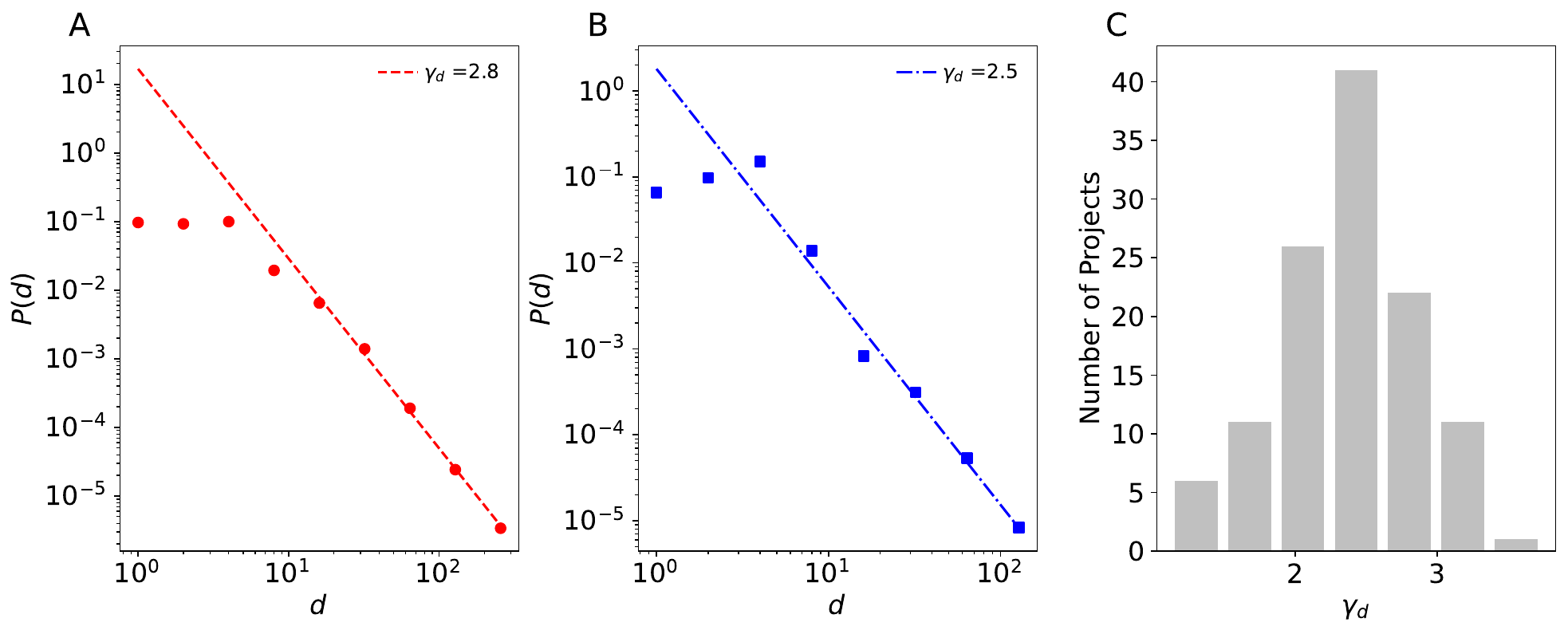}
\caption{Distribution of activity durations. A and B) The activity duration distributions for two construction project schedules (data points). The lines are fits to the power law tail $P(d)\sim d^{-\gamma_d}$. C) Histogram of the exponent $\gamma_d$ across construction project schedules.}
\label{fig_data_d}
\end{figure}

\section{Data for construction projects}

To investigate the statistics of activity durations in real projects I will use data for construction projects in the Nodes \& Links database. For each project we have a schedule that contains the duration and total float for every activity in the project. The total floats reported in those schedules were calculated by the planners making the schedules, using standard softwares such as Oracle Primavera P6 or Microsoft project. In total it amounted to 118 projects with 1,000 activities or more.

The inspection of the activity duration distribution for two construction project schedules reveals a power law scaling $P(d)\sim d^{-\gamma_d}$ for durations above 10 days (Fig. \ref{fig_data_d}A and B). The power law exponent $\gamma_d$ is project dependent and it is distributed around 2.5 (Fig. \ref{fig_data_d}C). Based on this data I conclude that activity durations are characterized by a power-law distribution for large durations. Note that the power-law distribution is scale-invariant or scale-free. When we change the duration scale from $d$ to $D = \alpha d$, the new duration distribution $P(D) = \alpha^{\gamma_d-1} D^{-\gamma_d}$ exhibits the same power law scaling.

\begin{figure}[t]
\includegraphics[width=6.3in]{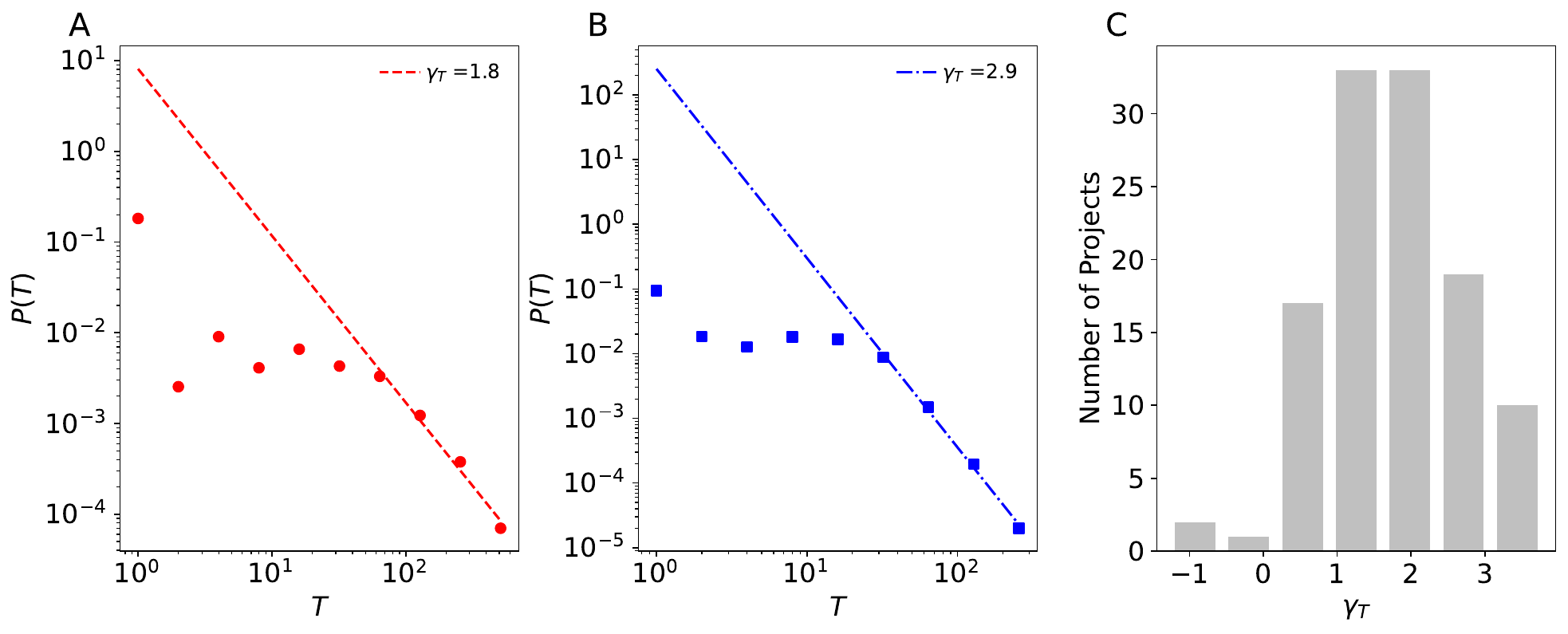}
\caption{Distribution of total floats. A and B) The activity total float distributions for two construction project schedules (data points). The lines are fits to the power law tail $P(T)\sim T^{-\gamma_T}$. C) Histogram of the exponent $\gamma_T$ across construction project schedules.}
\label{fig_data_T}
\end{figure}

The distribution of total floats $T$ across activities exhibits a power law tail $P(T)\sim T^{-\gamma_T}$ as well (Fig. \ref{fig_data_T}A and B). The power law exponent is project dependent and it is distributed around 1.5 (Fig. \ref{fig_data_T}C). As shown above, the activity durations are characterized by a power law distribution and the total floats are in some way related to activity durations. Based on this evidence we could rush to the conclusion that the scale-free nature of the total float distribution across activities is a consequence the associated distribution of durations. However, as shown below, this is not the case.

\section{Duplication-split model with duration dynamics}

Now I'll explain the observations in the previous section using the duplication split-model \cite{vazquez23}. The duplication-split model is driven by specialization. At specialization a parent activity is replaced by two new activities executing mutually exclusive components of the original activity. The simplest activity duration dynamics related to such specialization event is the halving of the activity duration: The two new activities inherit half the duration of the parent activity. This example
\begin{center}
\begin{tikzpicture}[->,>=stealth',auto,node distance=1.5cm,
	thick,main node/.style={font=\sffamily\small}]
\node[main node] (1) {1};
\node[main node] (2) [below left of=1] {1/2};
\node[main node] (3) [below right of=1] {1/2};
\node[main node] (4) [below left of=2] {1/4};
\node[main node] (5) [below right of=2] {1/4};
\node[main node] (6) [below left of=5] {1/8};
\node[main node] (7) [below right of=5] {1/8};
\path[every node/.style={font=\sffamily\small}]
	(1) 	edge (2)
		edge (3)	
	(2)    edge (4)
	        edge (5)
	(5)    edge (6)
	        edge (7);	      
\end{tikzpicture}
\label{example}
\end{center}
illustrates how the activity fragmentation process yields an activity durations distribution (1/2, 1/4, 1/8, 1/8) from a total duration 1.

The duplication-split model is the updated as follows.
\begin{enumerate}

\item Initial condition: Start with two activities A and B with duration $D_T/2$ and the dependency A$\rightarrow$B, where $D_T$ is the total time required to perform all project work.

\item Specialization step: Select one of the existing activities with equal probability, replace it by two new activities with half the duration of the parent and update the activity relations as follows:
\begin{enumerate}
\item {\em duplication}, with probability $q$ the new activities inherit all the predecessors and successors of the parent,
\item {\em split}, otherwise one new activity inherits all the predecessors and the other one all the successors of the parent  and a new arrow is created between the former and the latter.
\end{enumerate}

\item Repeat the specialization step until $N$ activities are created.

\end{enumerate}

\begin{figure}[t]
\includegraphics[width=6.3in]{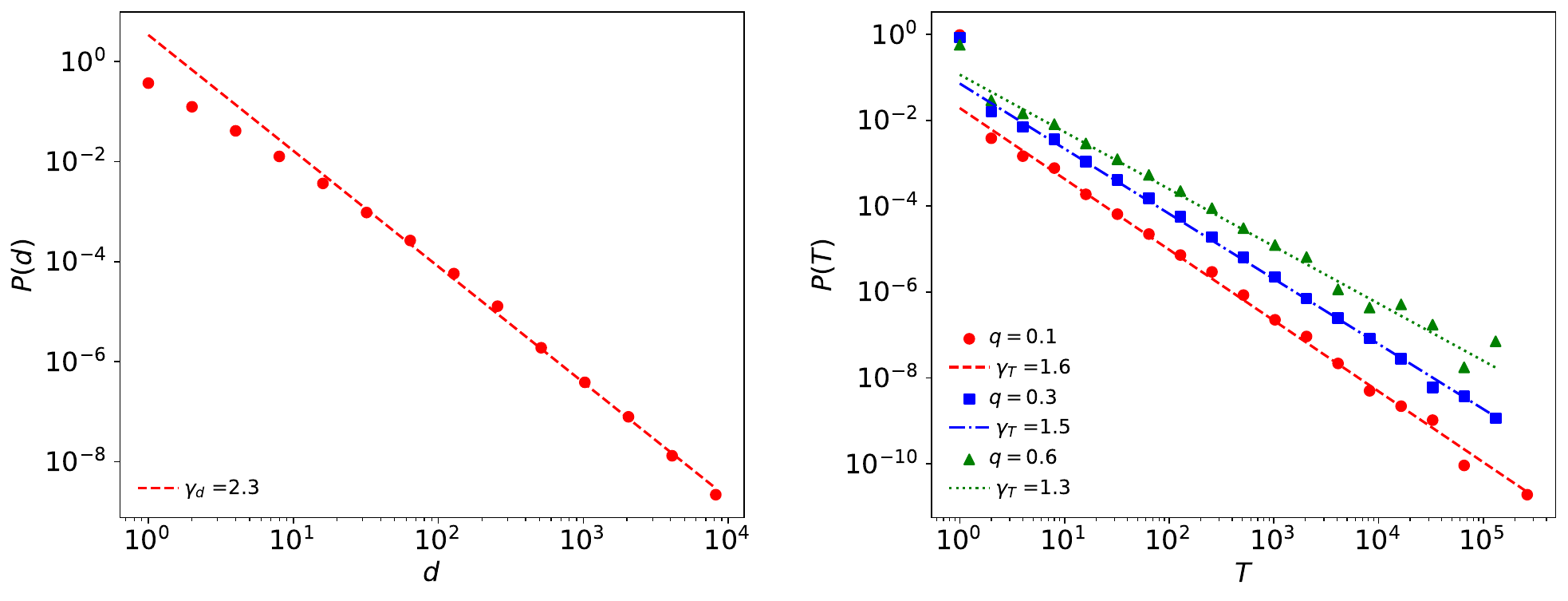}
\caption{Distribution of activity durations and total floats for the duplication-split model. A) Activity durations distribution. B) Activities total float distribution. The symbols are the result of numerical simulations of the duplication-split model. The lines are fits to the power law tails $P(d)\sim d^{-\gamma_d}$ and $P(T)\sim T^{-\gamma_T}$ respectively.}
\label{fig_model}
\end{figure}

I ran the duplication-split model for different values of $q$. I use a large size of $N=1$ million activities and I set $D_T=N$. The larger the $N$ values the more evident the power scalings of the activity durations and total floats.  Based on the duplication-split model rules, the activity duration statistics is independent of the duplication quotient $q$. Accordingly I report only one activity duration distribution. The distribution of activity durations follows a power law scaling $P(d)\sim d^{-\gamma_d}$ with $\gamma_d=2.3$ (Fig. \ref{fig_model}A). The sole process of halving activity durations yields the power law scaling. This does not come to a complete surprise. It has been reported that fragmentation can shatter a system into a collection of objects with scale-free size distributions  \cite{krapivsky94,sotolongo96}. The exponent value, $\gamma_d=2.3$ is around the most probable value observed for construction project schedules (Fig. \ref{fig_data_d}C).

Next, I use standard schedule methods to calculate the total floats using as input the network of activity dependencies and the activity durations \cite{van13}. (i) First we arrange the activities by topological order. Project schedules are acyclic: if we follow the arrows starting from one activity we will never comeback to that activity, regardless the activity we start from. We can order the activities such that for every dependency $A\rightarrow B$ the order of A is smaller than that of B. (ii) Next we execute a forward pass to calculate the early end dates $x_i$ of each activity $i$, where $i$ is the index over the topological order. We set $x_i=0$ for all activities. Following the topological order from $i=1$ to $i=N-1$ we update the early finish date as $x_j = \max( x_i  + d_j, x_j)$ if there is an arrow from $i$ to $j$, otherwise  $x_j = x_j$, for all $j>i$. That result in the project end date $x_N$. (iii) Finally, we perform a backward pass to calculate the late finish dates $y_i$. We set $y_i=x_N$ for all activities. Then, going from $i=N$ to $i=2$ we update the late finish dates according to the rule $y_j = \min(y_i - d_i, y_j)$ if there is an arrow from $j$ to $i$, otherwise $y_j = y_j$, for all $j<i$. The total float of each activity is the difference between its late and early finish dates $T_i = y_i - x_i$. 

The total floats are dependent on both the activity network and the activity durations. Therefore we expect differences in the total float distribution depending on the duplication quotient $q$. The distribution of total floats across activities follows a power law scaling $P(T)\sim T^{-\gamma_T}$, with an exponent $\gamma_T$ between 1 and 2 depending on the value of $q$ (Fig. \ref{fig_model}B). This range of exponents overlaps with the range where most projects lie (Fig. \ref{fig_data_T}C).

\begin{figure}[t]
\begin{center}
\includegraphics[width=3.3in]{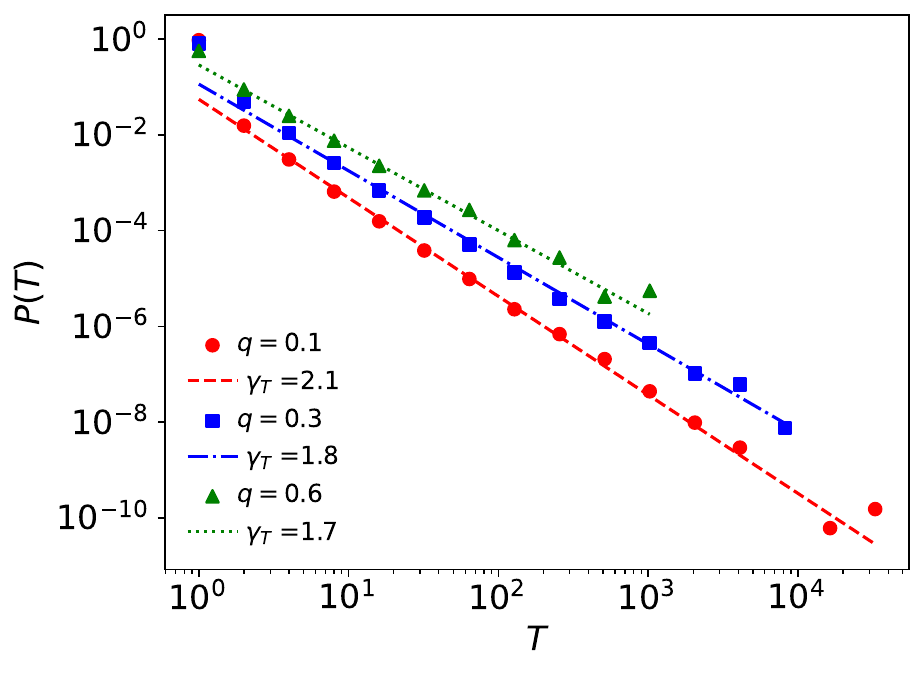}
\end{center}
\caption{Distribution of total floats for the duplication-split model with all activity durations equal to 1. The symbols are the result of numerical simulations of the duplication-split model. The lines are fits to the power law tail $P(T)\sim T^{-\gamma_T}$.}
\label{fig_model_d1}
\end{figure}

At this point we wonder what determines the power law scaling  of the total float distribution. The activity network, the activity durations, or both. To answer this question we repeated the total float calculations after setting all activity durations $d_i=1$. Surprisingly, it has the same power law scaling $p(T)\sim T^{-\gamma_T}$, albeit with different values of $\gamma_T$ (Fig. \ref{fig_model_d1}). From this analysis I conclude that the power law scaling $p(T)\sim T^{-\gamma_T}$ of the activities total float distribution is primarily a consequence of the activity network, but the exponent $\gamma_T$ is modulated by the activities duration distribution.

\section{Implications for delay risk analysis}

Projects do not proceed as planned. Delays in the activity durations often exceed the total floats, resulting in project completion delays \cite{ellinas19, park21}. Let us denote by $Z_i$ the increase in duration of the project activity with label $i$. I will call $Z_i$ the activity duration delay or simply the activity delay to abbreviate. In general the $Z_i$ are random variables with certain activity dependent cumulative distribution function $F_i(z) = {\rm Prob}(Z_i\leq z)$, denoting the probability
that the delay $Z_i$ of activity $i$ is less than $z$. Given a model of $F_i(z)$ for each activity,  we can estimate the cumulative distribution function $G(z) = {\rm Prob}(Z\leq z)$ that the project delay $Z$ is smaller than $z$ using the tropical approximation \cite{vazquez_tropical22}
\begin{equation}
G(z) = \prod_i  F_i(z + T_i).
\label{GFi}
\end{equation}
In a nutshell, if the project delay is less than $z$ then the delay of activity with label $i$ should be less than $z+T_i$.
Finally, the probability that the project delay is larger than $z$ is given by the tail distribution function
\begin{equation}
Q(z) = 1 - G(z).
\label{QG}
\end{equation} 

At this point we are tempted to introduce a PERT style uncertainty model, with triangular or beta distributions of activity durations (see for example Ref. \cite{hulett16}, Chapter 2). However, based on empirical data for construction projects, the statistics of activity delays is better represented  by a lognormal distribution ({\em The law of activity delays} \cite{vazquez23}). The simplest delay risk model has planned duration as a risk factor and the activity delays $z_i$ follows the log-normal cumulative distribution function \cite{vazquez23law}
\begin{equation}
F_i(z) = 1-p(\log(1+d_i)) + p(\log(1+d_i)) L\left(   \frac{\ln z - \mu_0 - \log(1+d_i )  }{ \sqrt{2}\sigma} \right).
\label{Fi}
\end{equation}
The term $p(\log(1+d_i))$ represents the delay likelihood and it is modulated by the logistic function
\begin{equation}
p(x) = \frac{1}{1+\exp[-g_0-g_1x]}.
\label{delay_likelihood}
\end{equation}
The term $L(\ldots)$ represents the delay impact and it is modulated by the cumulative distribution function of the standardized lognormal distribution 
\begin{equation}
L(x) = \frac{1}{2}[1 + {\rm erf}(x)],
\label{CDFlognormal}
\end{equation}
where ${\rm erf}(x)$ is the error function. The model parameters $g_0$, $g_1$, $\mu_0$ and $\sigma$ were estimated by fitting empirical data for activity delays. Their best fitting values are $g_0=-1.570$, $g_1=0.596$, $\mu_0=2.07$ and $\sigma=1.20$ (Ref. \cite{vazquez23law}, Table I, D model).

\begin{figure}[t]
\begin{center}
\includegraphics[width=6.3in]{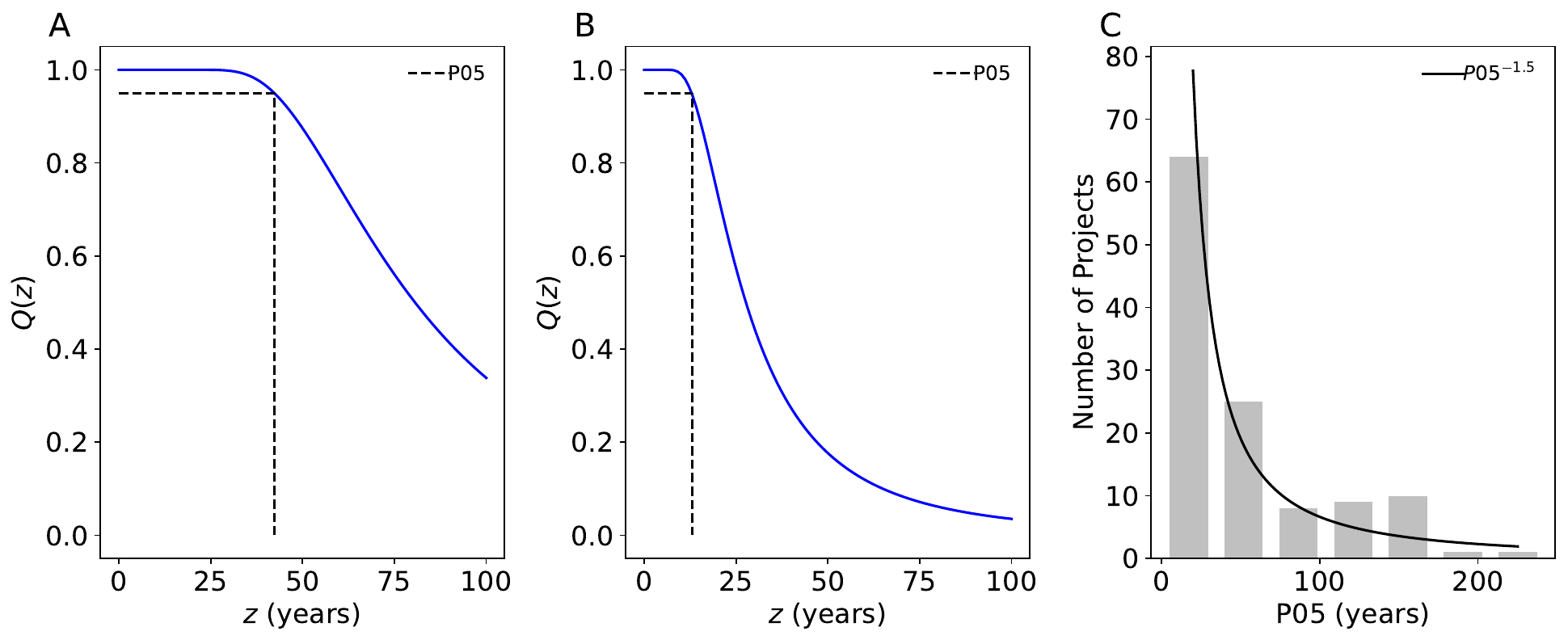}
\end{center}
\caption{A and B) Project delay distribution tail for two construction projects. The blue solid line is obtained using equation \ref{Qz}. The dashed line highlights the calculation of the P05. C) The distribution of P05s across construction projects (symbols). The solid line highlights the power law scaling.}
\label{fig_Qz_D}
\end{figure}

Combining equations (\ref{GFi})-(\ref{Fi}) we obtain the tail distribution for the project delay
\begin{equation}
Q(z) = 1 - \prod_i \left[ 1-p(\log(1+d_i)) +
 p(\log(1+d_i)) L\left(   \frac{\ln (z+T_i) - \mu_0 - \log(1+d_i )  }{ \sqrt{2}\sigma} \right) \right].
 \label{Qz}
\end{equation}
We can plug in actual data for activity durations and total floats into equation (\ref{Qz}) to get an estimate of the expected project delay tail distribution $Q(z)$. This is illustrated in Fig. \ref{fig_Qz_D}A and B (blue solid line) for two different projects, with $z$ in units of years. $Q(z)$ remains close to 1 for a few years, suggesting that there is no chance these two projects will finish on time. The dashed lines highlight the calculation of the P05, the delay value such that there is a 5\% chance that the project delay is smaller than that value. In other words, with 95\% confidence the project will finish after the P05. The P05s are of several years. Around 42 and 13 years for the projects shown in Fig. \ref{fig_Qz_D}A and B, respectively. Most projects have a P05 in the range of 10 years (Fig. \ref{fig_Qz_D}C), with some outliers with very large P05s.

\begin{figure}[t]
\begin{center}
\includegraphics[width=6.3in]{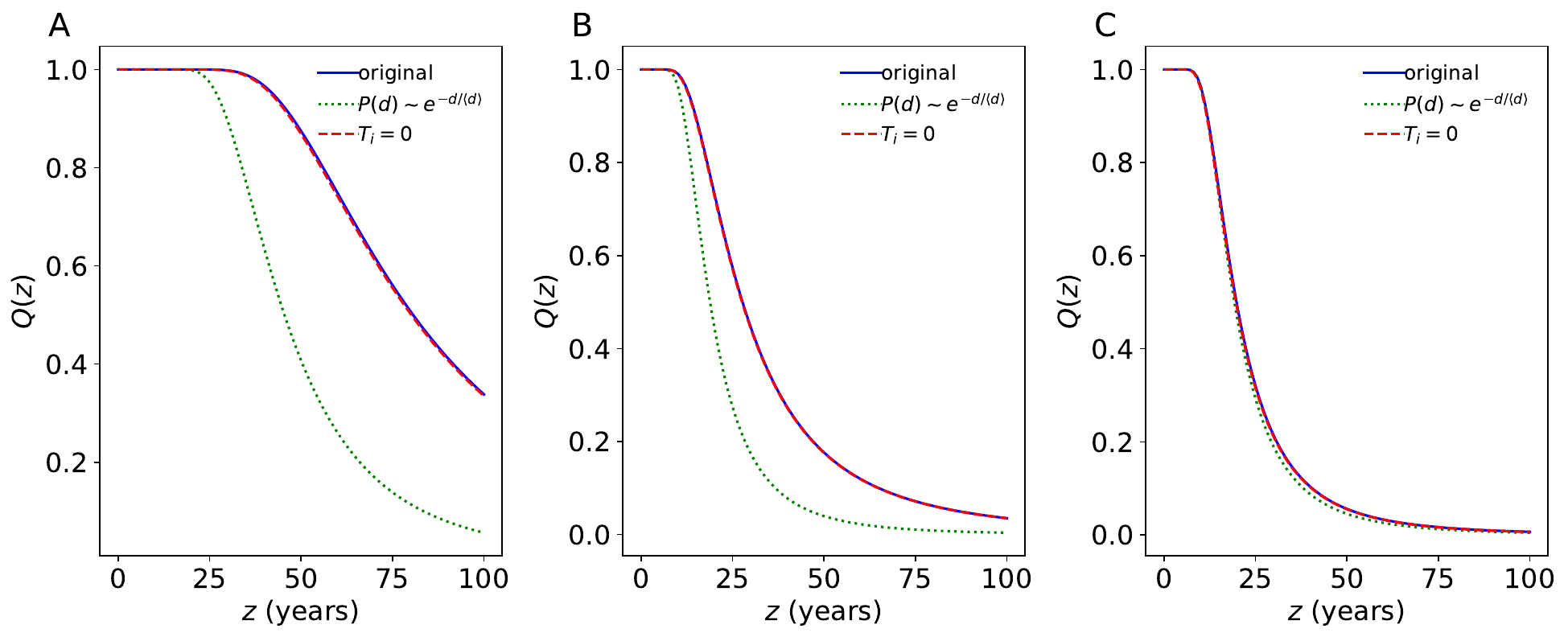}
\end{center}
\caption{Impact if the $P(d)$ and $P(T)$ scalings on the project delay tail distribution $Q(z)$ for 3 construction projects. The solid blue line was obtained using equation \ref{Qz} after plugging in the activity durations and total floats reported in the project schedules. The green dotted line was obtained after replacing the activity durations by random variables extracted from an exponential distribution with scale parameter $\langle d\rangle$, where $\langle d\rangle$ is the average activity duration for the project under consideration. The red dashed line was obtained after setting to zero the total floats of all activities.}
\label{fig_Qz_exp}
\end{figure}

\subsection{Relevance of $P(d)$ and $P(T)$}

We can decouple the contribution of $P(d)$ and $P(T)$ to the observed project delay distribution $Q(z)$ by by means of simulation. In the first simulation we replace the activity durations by random values extracted from an exponential distribution with scale parameter $\langle d\rangle$, where $\langle d\rangle$ denotes the average duration across activities for the project under consideration. After this transformation the recalculated $Q(z)$ is shifted to the left (Fig. \ref{fig_Qz_exp}, green dotted lines). The power law scaling of the activity durations distribution makes an important contribution to $Q(z)$. If we neglect that power law scaling we would underestimate the project delay. In contrast, the actual total float distribution seems to be irrelevant. When we set all activity total floats to 0, we observe no significant difference from the original $Q(z)$ (Fig. \ref{fig_Qz_exp}, red dashed line overlapping with the solid blue lines).

\section{On the shape of the project delay distribution}

At this point I would like to use the knowledge acquired above to make more general statements about the shape of the project delay tail distribution $Q(z)$. My starting point is the tropical approximation in equation \ref{GFi}. As explained in the original source (Ref. \cite{vazquez_tropical22}), this approximation follows from the sub-exponential nature of the activity delays distribution. The lognormal distribution of activity delays has a tail of the type $x^{-1}\exp( -log(x)^2)$ that decays slower than any exponential distribution as $x$ increases. Here, we have learned that we can set the total floats $T_i=0$ without any significant change to the project delay tail distribution $Q(z)$. Therefore, we can substitute equation \ref{GFi} by
\begin{equation}
G(z) \approx \prod_i  F_i(z).
\label{GFiT0}
\end{equation}
In the above equation we can group activities by their durations to obtain
\begin{equation}
G(z) \approx \prod_d  [F(z;d)]^{N(d)},
\label{GFiT0Pd}
\end{equation}
where $F(z;d)$ is the cumulative distribution function of activity delays, parametrized by the activity duration, and $N(d)=N\times P(d)$ is the number of activities with duration $d$. Note that $[F_d(z)]^{N(d)}$ is the cumulative distribution function of the maximum delay of $N(d)$ activities of duration $d$. If $N(d)$ is large and $z>0$, by the extreme value theorem \cite{resnick87}
\begin{equation}
 [F(z;d)]^{N(d)} \approx \exp\left\{ - \left[ \frac{z - \mu(N(d),d)}{\sigma(N(d),d)} \right]^{-\frac{1}{s}} \right\},
\label{Fzlim}
\end{equation}
where $s>0$, $\mu(N(d),d)$ and $\sigma(N(d),d)$ are the shape, location and scale parameters. Substituting the latter into equation (\ref{GFiT0Pd}) we obtain 
\begin{equation}
G(z) \approx \exp\left\{ - \sum_d \left[ \frac{z - \mu(N(d),d)}{\sigma(N(d),d)} \right]^{-\frac{1}{s}} \right\},
\label{Gzlim}
\end{equation}

The location parameter $\mu(N(d),d)$ sets the minimum project delay, while $\sigma(N(d),d)$ sets the typical project delay. Since we tend to measure activity durations and delays in units of days then $\mu(N(d),d) \sim 1$ day. In contrast, based on our numerical experiments above, typical project delays are in the order of years. Therefore, we can assume that
\begin{equation}
\mu(N(d),d)\ll \sigma(N(d),d).
\label{mu_sigma}
\end{equation}

To determine $\sigma(N(d),d)$, I use the following scaling argument. Suppose we have a project with $N(d)$ activities of duration $d$ and dependencies network $A_1\rightarrow A_2\rightarrow \cdots A_{N(d)}$. Then comes a new project manager that decides to lump together every two activities, resulting a project with $N(d)/2$ activities of duration $2d$ and dependencies network $A_{1+2}\rightarrow A_{3+4}\rightarrow \cdots A_{N(d)-1 + N(d)}$. Both projects are executing the same work and the typical project delay should be the same. We expect $\sigma(N(d),d) = \sigma(N(d)/2,2d)$ and by extrapolation  $\sigma(N(d),d) = \sigma(N(d)/r,rd)$, where $r$ is some renormalization factor. To satisfy this scaling constraint, the scale parameter should be of the form
\begin{equation}
\sigma(N(d),d) = \sigma(N(d) d).
\label{sigma}
\end{equation}

Based on these considerations (equations (\ref{mu_sigma}) and (\ref{sigma})), the project delay cumulative distribution in equation (\ref{Gzlim}) can rewritten as
\begin{equation}
G(z) \approx \exp \left[ - \left( \frac{z_c}{z} \right)^{ \frac{1}{s} } \right].
\label{GzAnsatz}
\end{equation}
where
\begin{equation}
z_c = \left\{ \sum_d [\sigma(N(d)d)]^{ \frac{1}{s} } \right\}^s,
\label{zc}
\end{equation}
is the typical project delay.

\subsection{Reference class forecasting}

Lumping together data for different projects is the basis of reference class forecasting (RFC) \cite{flyvbjerg06, natarajan22, flyvbjerg22, baerenbold23}, a methodology to estimate the statistics of some project quantity (delay, cost, etc) based on historical data. Here I claim that equation (\ref{GzAnsatz}) is the analytical demonstration of RFC when the quantity of interest is the project delay. I can say more. To apply equation (\ref{GzAnsatz}), and therefore RCF for project schedule overruns, we need to use the appropriate rescaling factor $z_c$. In other words, if we want to lump together project delay data for different projects we need to normalize them by $z_c$. 

The typical project delay parameter $z_c$ is of key importance when we lump together data from different projects. If we use the wrong $z_c$ then the resulting reference class distribution will not reflect the true project delay distribution. In all past implementations of RCF for project schedule the focus has been on the project duration $D$ and the normalization factor has been the planned project duration $D_0$ \cite{flyvbjerg06, natarajan22}. Moving from project duration overruns $D/D_0$ to normalized delays $(D-D_0)/D_0 = z/D_0$, we uncover that traditional RCF assumes $z_c = D_0$. However, there is no argument I am aware of that justifies that choice. We could use instead the total planned duration $D_T = \sum_i d_i$ or any other quantity that have units of durations and scales with the project size.

We need additional work to identify the right choice of $z_c$. Otherwise we would give wrong project duration estimates from historical data. The main guidance is the shape of the project delay cumulative delay distribution in equation (\ref{GzAnsatz}), with the unknown shape parameter $s$. We could try different choices of $z_c$ and check which one gives the best fit to equation (\ref{GzAnsatz}).

\subsection{The lognormal case}

An alternative approach to identify the correct $z_c$ is to deduce the functional dependency of $\sigma(n)$ in equation (\ref{zc}). The law of activity delays reported in Ref. \cite{vazquez23law} and discussed above, indicates that the activity delays follow a lognormal distribution. We are fortunate that mathematicians are ahead of us. They have lay the path to uncover $\sigma(n)$.

\begin{figure}[t]
\includegraphics[width=6.3in]{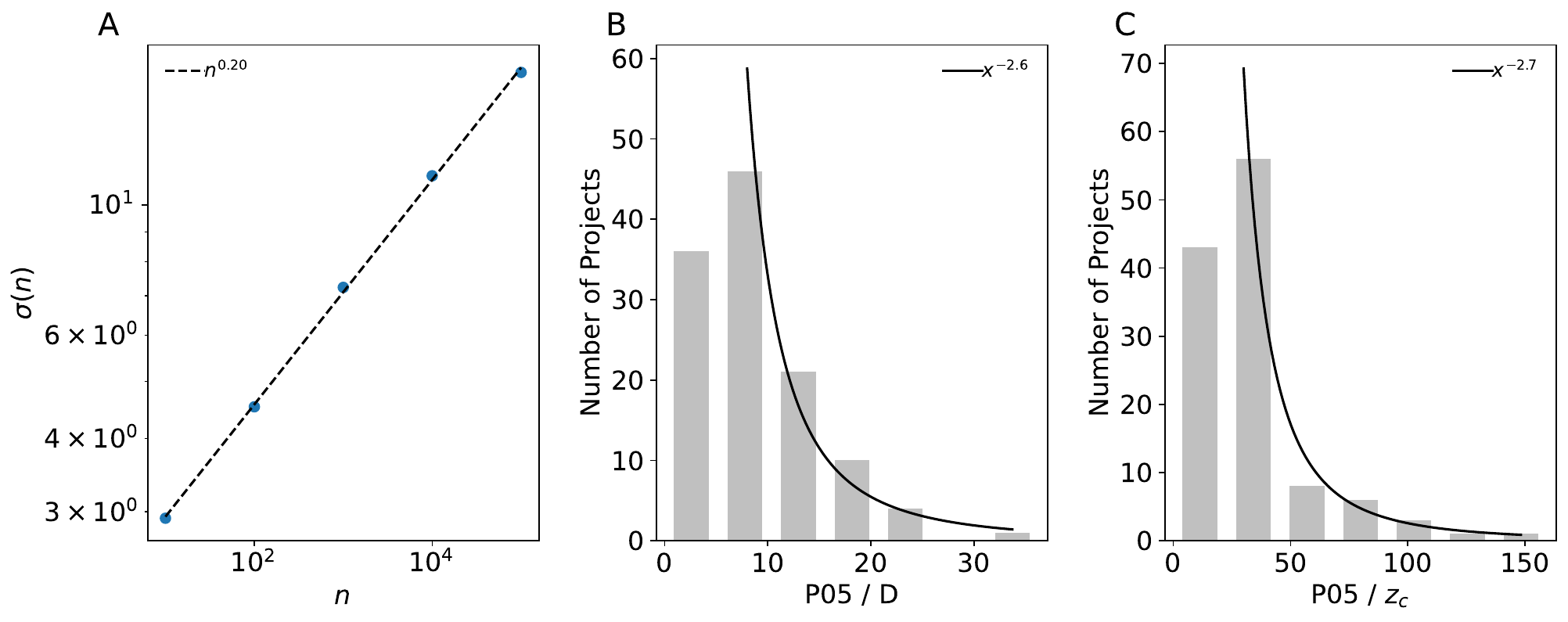}
\caption{A) Scaling of $\sigma(n)$ with $n$. The symbols are obtained from the numerical solution of equations (\ref{sigma_b}) and (\ref{bn}). The line is the best fit to a power law growth. B and C) Distribution of renormalized P05 using the project duration $D$ or $z_c = N^{0.20} (d_{\max}/d_{\min})^{1+0.20(1-\gamma_d)}$ as normalization factor. The lines are fits to a power law decay,.}
\label{fig_zc}
\end{figure}

If activities with the same duration are characterized by the same lognormal distribution, then $s=1$ and the normalization factor is (see Ref. \cite{liao12}, equation (1.1))
\begin{equation}
\sigma(n) = \frac{b(n)}{\log b(n)}
\label{sigma_b}
\end{equation}
where $b(n)$ is the solution of the equation
\begin{equation}
2\pi (\log b_n)^2 \exp\left( (\log b_n)^2\right) = n^2.
\label{bn}
\end{equation}
The numerical solution of these equations indicates that $\sigma(n)\approx 1.9 n^{0.20}$ (Fig. \ref{fig_zc}A). After substitution of this result into equation (\ref{zc}) we obtain
\begin{equation}
z_c = 1.9 \sum_d (N(d)d)^{0.20}.
\label{zc_lognormal}
\end{equation} 
If, as reported above, the activity durations distribution follows the power law scaling $P(d)\sim d^{-\gamma_d}$, then $N(d) \sim N d^{-\gamma_d}$, where $N$ is the number of activities. Substituting this result in the equation above
\begin{equation}
z_c \sim N^{0.20} \sum_d d^{0.20(1-\gamma_d)}.
\label{zc_lognormal_gamma_d}
\end{equation} 
Based on empirical data (Fig. \ref{fig_data_d}C) $\gamma_T$ is smaller than 4 and, therefore, $0.20(1-\gamma_d)<1$. In this case the sum in the latter equation is dominated by the largest values of $d$
\begin{equation}
\sum_d d^\alpha \sim \int^{d_{\max}} x^{0.20(1-\gamma_d)} dx \sim d_{\max}^{1+0.20(1-\gamma_d)},
\end{equation}
where $d_{\max}$ is the maximum activity duration. Plugging in this result into equation (\ref{zc_lognormal_gamma_d}) we finally obtain
\begin{equation}
z_c \sim N^{0.20} d_{\max}^{1+0.20(1-\gamma_d)}.
\label{zc_lognormal_gamma_d_dmax}
\end{equation} 

Now we go back to the P05 estimates reported in Fig. \ref{fig_Qz_D}C. In the absence of any normalization, the distribution of P05 across projects has a heavy tail. That heavy tail may be an artifact of lumping together projects with different gross properties. A traditional RCF approach is to normalize by the project duration $D$. We can inspect the distribution of P05 after normalization by $D$ in Fig.  \ref{fig_zc}B. This normalization does suppress the power law tail, increasing the power law exponent from 1.5 to 2.5. If instead we normalize by the $z_c$ deduced for the lognormal case (equation (\ref{zc_lognormal_gamma_d}), the power law exponent of the distribution increases further to value of 2.7. Another way to see the improvement by using $z_c$ over $D$ is to focus on the \% of projects with rescaled P05 above the most probable value. It takes the values 45\%, 30\% and 16\% in the case of no normalization, normalization by project duration $D$ and normalization by $z_c$, respectively. Therefore, the normalization by $z_c$ does a better job in suppressing the outliers than the normalization by $D$.

\section{Conclusions}

Based on data for construction project schedules, the distributions of duration and total float across activities are characterized by the power law scalings $P(d)\sim d^{-\gamma_d}$ and $P(T)\sim T^{-\gamma_T}$. The power law exponents $\gamma_d$ and $\gamma_T$ are project dependent.

The duplication-split model with activity duration dynamics recapitulates the power law scalings $P(d)\sim d^{-\gamma_d}$ and $P(T)\sim T^{-\gamma_T}$. We have previously shown that the duplication-split model reproduces the power law scalings of the distribution of in-coming and out-going dependencies \cite{vazquez23}. Taken together, this evidence indicates that the duplication-split model (i) captures the nature of the network and duration dynamics of project schedules and (ii) can be use as a generator for synthetic schedules to investigate other features.

Using the tropical approximation \cite{vazquez_tropical22} and the law of activity delays \cite{vazquez23law} we obtain an analytical equation for the project delay tail distribution, using as input the activity durations and total floats. By means of simulations, we demonstrated the importance of using the observed power law scaling of the activity duration distribution. In contrast, the power law scaling of the total float distribution appears to be irrelevant  for the distribution of project delay.

Finally, using the extreme value theory and some scaling arguments, we derived a general equation for the project delay distribution. This equation represents a mathematical proof for RFC of project delay.
From this equation and the assumption of a lognormal distribution of activity delays, as suggested by the law of activity delays \cite{vazquez23law}, we obtain a typical project delay of the order of $z_c \sim N^{0.20}d_{\max}^{1+0.20(1-\gamma_d)}$, where $N$ is the number of activities, $d_{\max}$ the maximum activity duration in units of days and $\gamma_d$ the power law exponent of the activity duration distribution. Based on numerical experiments $z_c$ does a better job than the project duration as a normalization factor when aggregating project delays from different projects. Further work is required to asses the use of $z_c$ as a normalization factor for project delays using historical data.

\section*{Acknowledgements}

I thank the data science team at Nodes \& Links, in particular Hidayet Zaimaga and Georgios Kalogridis, for the generation of the construction project schedules database. 
{\bf Funding:} This research did not receive any specific grant from funding agencies in the public, commercial, or not-for-profit sectors. Nodes \& Links Ltd provided support in the form of salary for A.V., but did not have any additional role in the conceptualization of the study, analysis, decision to publish, or preparation of the manuscript.
{\bf Author contributions:} A.V. conceptualized and executed the work.
{\bf Competing interests:} The author declares that he has no competing interests.
{\bf Data and materials availability:} The data for real schedules is fully represented in the figures. The code for the duplication-split model is available at \href{https://github.com/av2atgh/red}{github.com/av2atgh/red}.

\bibliographystyle{unsrt}

%\bibliography{risk.bib}

\end{document}